\documentclass[iop]{emulateapj}
\usepackage{graphicx}
\usepackage{natbib}
\usepackage{latexsym}
\usepackage{amssymb}
\usepackage{hyperref}

\makeindex

\begin{document}
\submitted{Accepted for publication in ApJ, 2015 June 8}

\title{HST/WFC3 Observations of an Off-Nuclear Superbubble in Arp 220}

\author{Kelly E. Lockhart\footnotemark[1], 
Lisa J. Kewley\footnotemark[2],
Jessica R. Lu\footnotemark[1],
Mark G. Allen\footnotemark[3],
David Rupke\footnotemark[4],
Daniela Calzetti\footnotemark[5],
Richard I. Davies\footnotemark[6],
Michael A. Dopita\footnotemark[2],
Hauke Engel\footnotemark[6],
Timothy M. Heckman\footnotemark[7],
Claus Leitherer\footnotemark[8],
David B. Sanders\footnotemark[1]
}

\footnotetext[1]{Institute for Astronomy, 2680 Woodlawn Drive, Honolulu, HI 96822, USA}
\footnotetext[2]{Research School of Astronomy and Astrophysics, Australian National University, Cotter Road, Weston Creek ACT 2611, Australia}
\footnotetext[3]{Observatoire de Strasbourg, UMR 7550, Strasbourg 67000, France}
\footnotetext[4]{Department of Physics, Rhodes College, Memphis, TN 38112, USA}
\footnotetext[5]{Department of Astronomy, University of Massachusetts, Amherst, MA 01003, USA}
\footnotetext[6]{Max-Planck-Institut f{\"u}r Extraterrestrische Physik, Postfach 1312, 85741 Garching, Germany}
\footnotetext[7]{Center for Astrophysical Sciences, Department of Physics and Astronomy, Johns Hopkins University, Baltimore, MD 21218, USA}
\footnotetext[8]{Space Telescope Science Institute, 3700 San Martin Drive, Baltimore, MD 21218, USA}

\begin{abstract}
We present a high spatial resolution optical and infrared study of the circumnuclear region in Arp 220, a late-stage galaxy merger. Narrowband imaging using HST/WFC3 has resolved the previously observed peak in H$\alpha$+[NII] emission into a bubble-shaped feature. This feature measures 1.6'' in diameter, or 600 pc, and is only 1'' northwest of the western nucleus. The bubble is aligned with the western nucleus and the large-scale outflow axis seen in X-rays. We explore several possibilities for the bubble origin, including a jet or outflow from a hidden active galactic nucleus (AGN), outflows from high levels of star formation within the few hundred pc nuclear gas disk, or an ultraluminous X-ray source. An obscured AGN or high levels of star formation within the inner $\sim$100 pc of the nuclei are favored based on the alignment of the bubble and energetics arguments.
\end{abstract}

\section{INTRODUCTION}

Galaxy mergers are integral to understanding galaxy evolution. Simulations suggest that major mergers cause large-scale tidal gas inflows, which may fuel nuclear starbursts and active galactic nucleus (AGN) activity \citep{barnes1996transformations, mihos1996gasdynamics}. In later stages, mergers may host young, obscured AGN, which, simulations predict, are gradually uncovered as AGN feedback and high levels of circumnuclear star formation expel the surrounding gas \citep{barnes1996transformations, mihos1996gasdynamics, di-matteo2005energy, hopkins2005black, hopkins2006fueling}. This nuclear activity can, in turn, fuel large-scale galactic outflows \citep{heckman2003starburst-driven, veilleux2005galactic}, which have been observed in many merging systems \citep{heckman1990on-the-nature, heckman2000absorption-line, martin2005mapping, rupke2002keck, rupke2005outflows3, rupke2005outflows1, rupke2005outflows2, martin2006mapping, rich2010ngc-839:, rupke2011integral, rupke2013the-multiphase}. 

Large-scale outflows in the outer parts of galaxies have been extensively studied. However, the advent of high spatial resolution observations has prompted more recent studies to probe the launch sites, propagation, and escape of these outflows \citep{rupke2013breaking, emonts2014outflow, sakamoto2014an-infrared-luminous}. When observed with sufficient spatial resolution, an outflow still within a galaxy often has complex superbubble or filamentary structure \citep{veilleux1994the-nuclear, forbes2000a-multiwavelength, cecil2001jet-, kenney2002hubble}, created as warmer, high-velocity material ejected from a galaxy's nuclear region interacts with the cooler interstellar medium (ISM). However, these well-shaped structures have thus far not been observed in the turbulent, clumpy ISM of a late-stage merger. In addition, while both the timescale and impact of AGN feedback into the host system have been calculated theoretically \citep{hopkins2005black, hopkins2010quasar}, both are still observationally uncertain. Investigating an individual late-stage merging system that exhibits both an outflow and a possible obscured AGN may provide clues to this process.

\begin{figure*}
\begin{center}
\includegraphics[width=\textwidth]{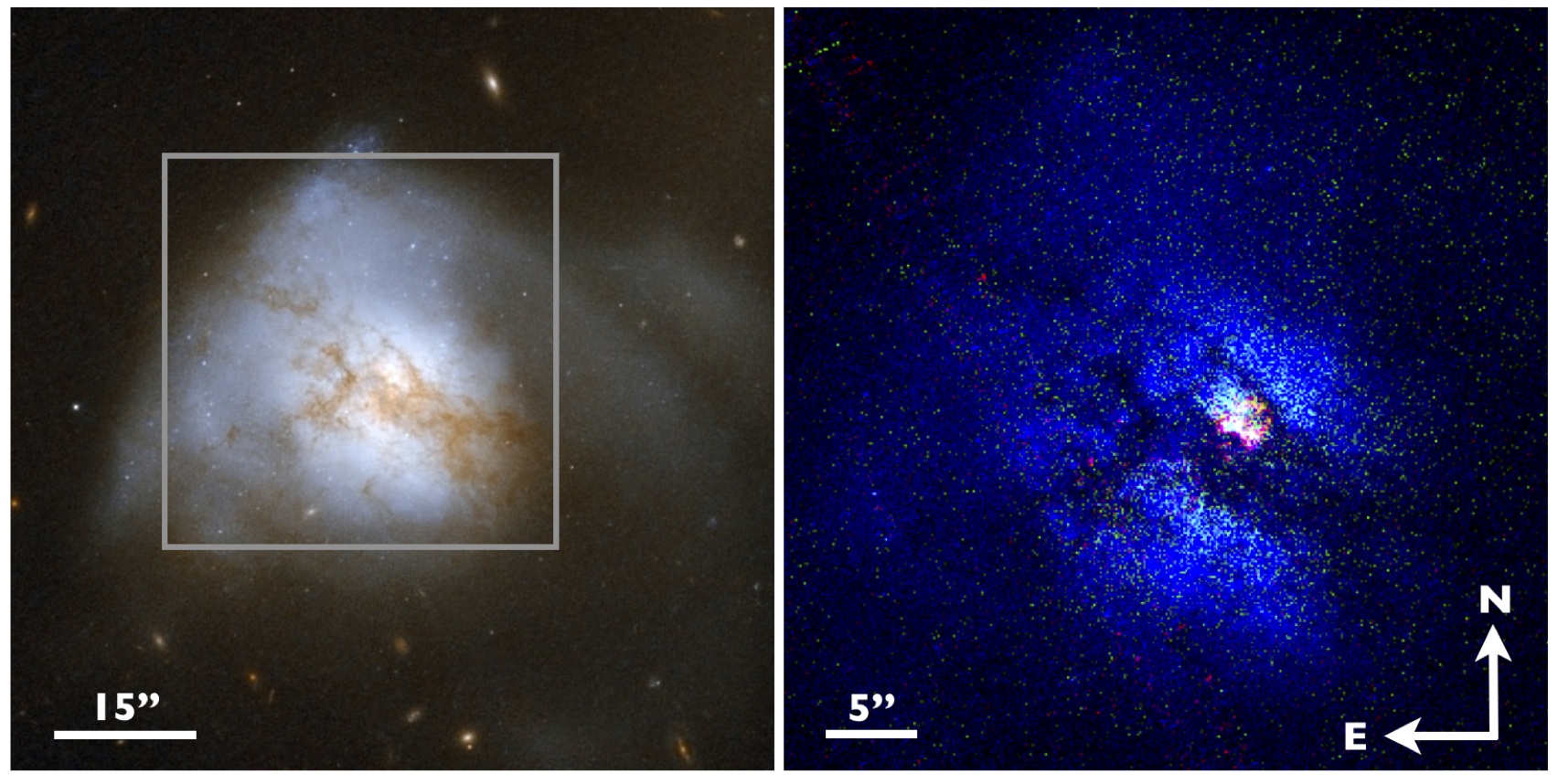} 
\caption{\emph{Left:} \textit{HST} ACS image of Arp 220 as first presented in \citet{wilson2006two-populations}, rotated to adjust for the 77$^{\circ}$ position angle. The blue image is taken with the F435W filter and the red with the F814W filter (green in a linear combination of the red and blue filters). The gray box indicates the FOV of the zoomed right panel. \emph{Right:} Our new data taken with \textit{HST} WFC3. Blue is the F467M blue continuum image, green is [OIII], or the FQ508N image with the blue continuum subtracted, and red is H$\alpha$+[NII], or the F665N image with the red continuum subtracted. North is up, east is left.} 
\label{fig:HSTacs}
\end{center}
\end{figure*}

Arp 220 is the closest ultraluminous infrared galaxy \citep[ULIRG, L$_{IR}$ \textgreater 10$^{12}$ L$_{\odot}$,][]{sanders1988ultraluminous, soifer1987the-iras}. The system has long been recognized as a late-stage merger, as evidenced by extended tidal tails visible in the optical \citep{arp1966atlas, surace2000high-resolution}, infrared \citep{joseph1985recent}, and in HI \citep{hibbard2000the-neutral}, as shown in figure~\ref{fig:HSTacs}. It also hosts a large--scale outflow that extends far from the galaxy. Arp 220 has a molecular gas mass of $\sim$10$^{10}$ M$_{\odot}$ \citep{scoville1997arcsecond} and a total star formation rate, calculated using the far-infrared luminosity, of 240$\pm$30 M$_{\odot}$ yr$^{-1}$ \citep{farrah2003starburst}. Most of the molecular gas and star formation are concentrated within a central few hundred pc nuclear gas disk, which also hosts a double nucleus. A possible AGN has been suggested \citep{paggi2013two-compton-thick} but is still debated. At a distance of 77 Mpc, its close proximity offers a unique laboratory to investigate the interaction of its outflow with the ISM at small spatial scales, and to determine how energy propagates out of the interior of such a system. A more detailed background on the structure of Arp 220 is presented below. 

\citet{heckman1987evidence} observed large-scale (10 kpc) bipolar lobes extending from the galaxy via narrowband H$\alpha$ imaging, which they interpret as a superwind. Further spectroscopic study revealed these lobes to be superwind-driven bipolar bubbles \citep{heckman1990on-the-nature}. This superwind is also observed in soft X-ray emission \citep{mcdowell2003chandra}. 

In addition to the large-scale wind, smaller-scale gas dynamics have been investigated in more recent optical IFU studies.  \citet{colina2004integral} found large-scale plumes and lobes, up to 8 kpc in length and aligned with the superwind, which are consistent with high-velocity shocks expanding in a neutral medium. \citet{arribas2001two-dimensional} studied the inner region of the system within 2 kpc of the nuclei and found three gas components. One component is coupled to the central molecular gas disk. The other two components are associated with a biconal outflow or superwind. 

This outflow, whether star formation- or AGN-driven, almost certainly originates within the nuclear region of the galaxy. At the center of Arp 220, two dust-enshrouded nuclei, separated by 0.98'' (368 pc), are visible in the IR \citep{scoville1998nicmos}, sub-mm \citep{scoville1997arcsecond, downes1998rotating, sakamoto1999counterrotating, scoville2014alma}, and radio \citep{baan1995nuclear}. \citet{sakamoto1999counterrotating} resolved the large circumnuclear gas mass in a detailed CO study, and found that each nucleus is embedded in its own gas disk, each roughly 100 pc in radius, and these disks are counterrotating with respect to each other and inclined such that their near sides are to the south. The position angle (P.A.) of each of these small disks is 52$^{\circ}$ and 263$^{\circ}$ for the eastern and western nuclei, respectively, while that of the larger circumnuclear disk is 25$^{\circ}$. \citet{scoville2014alma} calculated the interstellar medium (ISM) masses of each nuclear gas disk, finding a mass of 1.9 (4.2) $\times$10$^9$ M$_{\odot}$ around the eastern (western) nucleus, concentrated within a radius 69 (65) pc of the nucleus. These small gas disks are surrounded by a larger gas disk $\sim$1 kpc in radius, which rotates around the dynamical center of the system \citep{sakamoto1999counterrotating}. \citet{mundell2001nuclear} used observations of HI in the circumnuclear gas disks and comparison with simulations to estimate that the merger started about 700 Myr ago, with the second, most recent starburst starting in the western nuclear gas disk more recently than 10--100 Myr ago.

Within the inner 50 pc of each nucleus, the average molecular gas density is $\sim$10$^5$ cm$^{-3}$ \citep{scoville2014alma}, comparable to the conditions in a dense star-forming core inside a giant molecular cloud. Indeed, Arp 220 is undergoing very high rates of star formation. \citet{smith1998a-starburst} estimate the circumnuclear star formation rate at 50--800 M$_{\odot}$ yr$^{-1}$ using the radio supernovae rate, while \citet{farrah2003starburst} use the far-infrared luminosity to estimate a global star formation rate of 240$\pm$30 M$_{\odot}$ yr$^{-1}$.

Outside of the heavily extincted region near the nuclei, star formation activity was observed with \textit{HST} ACS. \citet{wilson2006two-populations} found two populations of massive star clusters: young ($<$10 Myr) massive clusters within 2.3 kpc of the nuclei and intermediate age ($\sim$100 Myr) clusters beyond this radius. 

In this paper, we investigate the connection between the nuclear energy sources and the large-scale outflow. We use HST/WFC3 with narrowband filters to create the first high spatial resolution optical emission line maps of Arp 220. We present maps of [OIII], H$\beta$, and H$\alpha$+[NII] to resolve gas motions in Arp 220. We discover a bubble feature in H$\alpha$+[NII] (section~\ref{sec:morph}) and discuss its origins in section~\ref{sec:disc}.

Throughout this work, we assume a distance of 77 Mpc, which gives 1''=375 pc.

\section{OBSERVATIONS}\label{sec:obs}

\begin{deluxetable*}{lcclcccc}
\tablecaption{HST WFC3-UVIS Exposures}
\tablewidth{0pt}
\tablehead{
\colhead{} & \multicolumn{2}{c}{Bandpass} & \colhead{} & \colhead{} & \multicolumn{2}{c}{Continuum} & \colhead{Flux Error} \\ 
\colhead{Filter}  & \colhead{Central $\lambda$ (\AA)} & \colhead{Width (\AA)} & \colhead{Line} & \colhead{Exposures} & \colhead{Filter}  & \colhead{Scale$^a$} & \colhead{1$\sigma^b$ (erg s$^{-1}$ cm$^{-2}$ \AA$^{-1}$ arcsec$^{-2}$)}
}
\startdata
FQ492N & 4933 & 114 & H$\beta$ & 2$\times$1090 s, 6$\times$920 s & F467M & 0.42 & 3.9 $\times$ 10$^{-18}$ \\
FQ508N & 5091 &131 & [OIII] 4959, 5007 & 2$\times$1050 s, 6$\times$920 s & F467M & 0.46 & 3.8 $\times$ 10$^{-18}$\\
F665N & 6656 & 131 & H$\alpha$ + [NII] 6548, 6583 & 2$\times$200 s & F621M & 0.16 & 1.4 $\times$ 10$^{-17}$\\
F680N & 6877 & 371 & [SII]$^c$ 6717, 6731 & 2$\times$525 s & F621M & 0.66 & 7.7 $\times$ 10$^{-18}$\\
F467M & 4683 & 201 & blue continuum & 2$\times$775 s & ... & ... & ... \\
F621M & 6219 & 609 & red continuum & 2$\times$155 s & ... & ... & ... 
\enddata
\tablenotetext{a}{Scaling factor of the continuum used for continuum subtraction.}
\tablenotetext{b}{Flux errors are estimated based on the dispersion of the sky background in each image, as measured on a section of sky away from the galaxy.}
\tablenotetext{c}{Not used for analysis due to shallowness of image; see Appendix}
\label{tab:obs}
\end{deluxetable*}

\subsection{HST}

\begin{figure}[htbp]
\begin{center}
\includegraphics[width=\columnwidth]{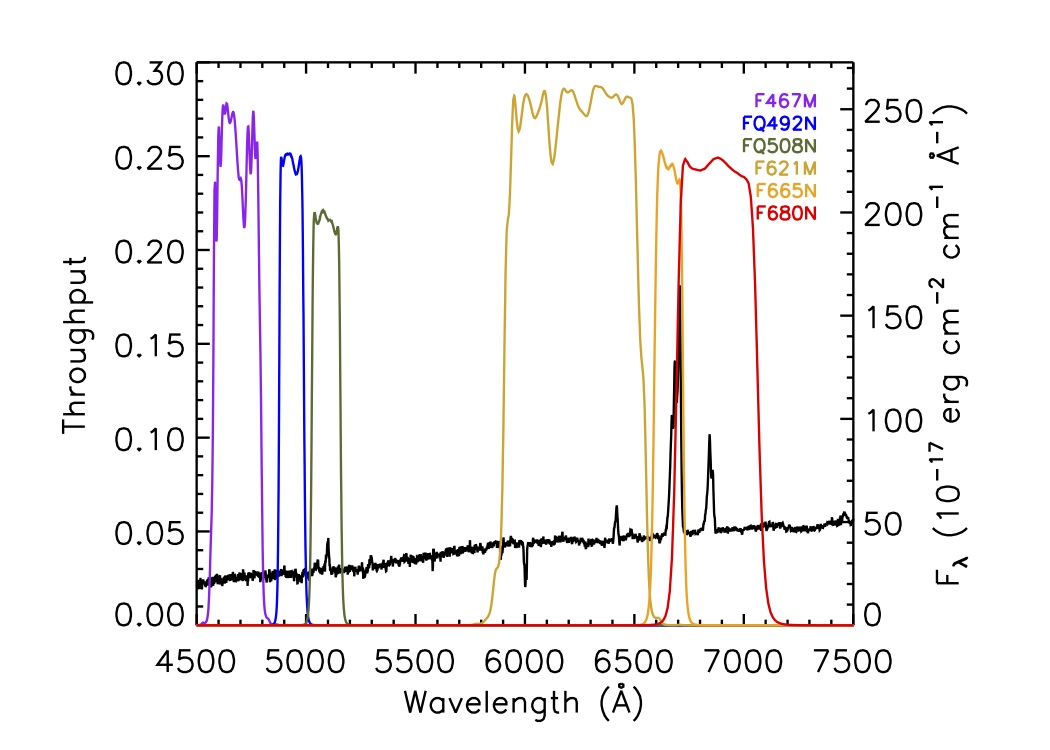}
\caption{SDSS 7 spectrum of Arp 220 with the WFC3-UVIS filter bandpasses overlaid. The left axis shows the throughput of each WFC3 filter (colors are as shown in the legend) while the right axis shows the flux of the SDSS spectrum (in black).}
\label{fig:SDSS}
\end{center}
\end{figure}

The optical imaging data of Arp 220 used in this work are new observations obtained with HST Wide Field Camera 3 (WFC3-UVIS) on 2012 September 22 (GO-12552, PI: Lisa Kewley). We selected filters whose bandpasses each cover one or more of several strong optical emission lines commonly used for calculating gas-phase metallicity and shock excitation: H$\beta$ (filter FQ492N); [OIII]$\lambda\lambda$ 4959, 5007 (rest) (FQ508N); H$\alpha$ + [NII]$\lambda\lambda$ 6548, 6583 (F665N); and [SII]$\lambda\lambda$ 6717, 6731 (F680N). Blue (F467M) and red (F621M) continuum exposures were also taken to allow calculation of the line flux across the extent of the system. Figure~\ref{fig:SDSS} shows the filter bandpasses superimposed over the SDSS 7 spectrum of Arp 220. The filter and exposure data are summarized in Table~\ref{tab:obs} along with the continuum bandpass used with each emission line image and the 1$\sigma$ RMS errors (discussed further in sections~\ref{ssec:csub} and \ref{ssec:err}) With a pixel size of 0.0396'', the field of view (FOV) of WFC3-UVIS is 164'' $\times$ 176'' as measured on the image. The FQ, or quad, filters, each occupy one-quarter of the FOV, so the FOV for these filters is approximately 82'' $\times$ 88''.

Data reduction was performed using the STScI pipeline, which utilizes ``On-The-Fly Reprocessing'' (OTFR). This processes the raw data files using the best calibration files and software available at the time of retrieval from the archive and performs basic bias and dark subtraction and flat fielding. It also flags bad pixels and performs initial cosmic ray removal. The data were all retrieved from the archive in late 2012.

The reduced dithered exposures were combined using AstroDrizzle \citep{avila2012astrodrizzle:}. AstroDrizzle, the successor to MultiDrizzle, provides distortion corrections and combines multiple images using a technique which allows for image subsampling. Sky subtraction was performed using the mode of the image, in order to not bias the sky value by the extended nebulosity of the galaxy. Charge transfer efficiency (CTE) trails caused by inefficient electron shuffling during readout were evident on the initial combined images. Due to the high counts involved, they were especially problematic in the pixels along which cosmic ray hits were shuffled. The charge in the first smeared pixel amounted to roughly 1\% of the initial high pixel value and dropped off rapidly as the charge was further shuffled along the detector. However, the CTE trail left by each cosmic ray was up to 20 pixels long for the strongest cosmic rays. These pixels were masked in AstroDrizzle, as the data were downloaded and reduced before the CTE correction software from STScI became available. All other AstroDrizzle parameters were set to the default values. The final combined images were registered to a common reference frame using the AstroDrizzle tasks \texttt{tweakreg} and \texttt{tweakback} and redrizzled onto the same pixel grid, to allow for image combination. 

The FQ492N and FQ508N filters, covering H$\beta$ and [OIII], had a sufficient number of exposures to allow for cosmic ray removal (see table~\ref{tab:obs}). The other filters (F467M, F621M, F665N, F680N) had two dithered exposures each; the cosmic ray cleaning provided by AstroDrizzle for these was not robust, especially in the gap between the two detector chips. The final combined images for these filters were further cleaned of cosmic rays using a Laplacian edge detection algorithm, which identifies cosmic rays based on the sharpness of their edges \citep{van-dokkum2001cosmic-ray}. This method, which removes only the positive noise spikes, can alter the noise distribution of the image. The resulting noise distribution may be asymmetric and skewed towards negative values, which can be problematic for low signal-to-noise ratio (S/N) regions. We examined the cleaned images and found that only the F665N image is affected. Because we focus on the high S/N regions in this filter, this does not affect our results.

The bandpass of one of the continuum images, through filter F621M, contains [OI]$\lambda$ 6300\AA, a strong emission line in shock-excited gas. The strength of this line varies with metallicity and excitation source and thus could contaminate calculations of emission line ratios. The contribution of the flux of this line to the total flux in the continuum image was estimated using the Sloan Digital Sky Survey (SDSS) spectrum of Arp 220. The SDSS fiber covers the inner 3'' of the galaxy, or roughly 1 kpc. The ratio of flux in the emission line to flux in the continuum contained in the bandpass was roughly 1\%, or less than the average error in the images. Thus, the contribution of the line to the continuum image was taken to be negligible and no correction to the continuum image was made.

The data were binned in square bins 3 pixels wide to increase the S/N. At the redshift of 0.018126$\pm$0.000023 \citep{de-vaucouleurs1991third}, the WFC3 pixel scale of 0.0396''/pixel corresponds to 15 pc. Extragalactic giant HII regions range from 50 to 300 pc in diameter \citep{oey2003h-ii-regions}. The binned resolution elements measure 45 pc across and thus allow full sampling of all but the smallest HII regions while still allowing recovery of low signal.

\subsubsection{Continuum Subtraction}\label{ssec:csub}

One of the most sensitive calibration tasks is subtraction of the stellar continuum from the emission line images. Small changes in the factor used to scale the continuum before subtracting it from the emission line image propagate into a change in the flux of the final emission line image. This can have a large effect on the emission line flux and ratio maps. 

We determined the optimal scale factor between the continuum image and the emission line image using the ``skewness transition method'' \citep{hong2014quantitative}. This method uses the symmetry of the flux distribution within an aperture to identify the ideal scaling factor that neither under nor over subtracts the continuum. It accounts for spatially dependent luminosity changes, but not spatially dependent color changes. The location of the aperture used to calculate the scaling factor influences the derived value, and allows us to estimate an error on the scaling factor. The error on the scaling factor can be taken as 0.05 for each filter.

We verified the accuracy of this method via physical arguments. For a star cluster with zero extinction, and assuming Case B recombination, the ratio of H$\alpha$ to H$\beta$ is 2.86. However, Arp 220 has high levels of patchy extinction and a high and varying ratio of [NII] to H$\alpha$, so we would not expect this ratio to hold across most of this system. We are able to check the ratio between the two images using a star cluster with significant H$\alpha$+[NII] and H$\beta$ emission, and a measured E(B-V) of 0 \citep[their cluster 62]{wilson2006two-populations}. We assumed that the ratio of [NII] to H$\alpha$ is roughly that for an HII region \citep[$\sim$0.3,][]{kewley2006metallicity} and adjusted the scaling factor for the continuum subtraction from the H$\beta$ image until we achieve a ratio equal to the Balmer decrement. With this method, we calculate a new scaling factor of 0.49$\pm$0.02. This scaling factor is generally consistent with that obtained through the skewness transition method (0.42$\pm$0.05), and so we adopt the scaling factors from the skewness transition method. Our scaling factors are given in table~\ref{tab:obs}.

\subsubsection{Error estimation}\label{ssec:err}

Flux errors are estimated based on the standard deviation of the background in each image. The dispersion varies slightly across the field, so to obtain a representative value this was measured in a region centered within 2'' of the detected emission line flux in each filter, but not overlapping it. We show the 1$\sigma$ errors in Table~\ref{tab:obs}. We adopt a S/N cutoff of 4$\sigma$ in all emission line images. However, the F680N image was too shallow and we were unable to recover [SII] flux above the 4$\sigma$ limit (see Appendix). We thus drop this image from further analysis.

\subsection{Other data}
\subsubsection{H$_2$ data}
Near-infrared (H and K band) integral field spectroscopy (IFS) data were taken 2007 March 7 and April 18--21 with SINFONI, a NIR integral field spectrometer on the Very Large Telescope (VLT) \citep{eisenhauer2003sinfoni}. The curvature-based adaptive optics system was operated in laser guide star mode, though without tip-tilt correction as no suitable tip-tilt star was available. This gave a nearly symmetric Gaussian point spread function of 0.30'' $\times$ 0.31''. The 0.05''$\times$0.10'' plate scale was used, giving a FOV of 3.2''$\times$3.2''. Further details on the observations, data reduction, and line fitting are given in \citet{engel2011arp-220:}.

\subsubsection{NICMOS}

Near-infrared imaging using HST/NICMOS was obtained from the HST archives. The F160W filter on camera NIC2 was used to image Arp 220 in the H-band on 10 January 2004 (proposal 9726, PI: R. Maiolino) for a total exposure time of 600 s \citep{cresci2007a-nicmos}. The F222M filter (K band) on NIC2 was used to image the system on 4 April 1997 for a total exposure time of 1024 s \citep[proposal 7116, PI: N. Scoville]{scoville1998nicmos}. The pixel scale on the NIC2 camera is 0.076'' while the FOV is 19.2''.

\subsection{Registration}
\label{ssec:reg}

There are no stars suitable for image registration in the field of view (FOV). Instead, bright super star clusters from \citet{wilson2006two-populations} were used to align the WFC3 and NICMOS datasets. Three bright, well-separated clusters were used, allowing a correction for both shift and rotation between the images. The absolute astrometric registration was set by the coordinates of the western nuclei given by \citet{scoville1998nicmos}, precessed to J2000.0.

The SINFONI data have a smaller FOV, limited to the circumnuclear region. The optical super star clusters (SSCs) previously used for alignment were out of this FOV. Instead, the IFU data were aligned directly with the NICMOS F222M image using the nuclei and another bright infrared source.

\begin{figure*}[t]
\begin{center}
\includegraphics[width=\textwidth]{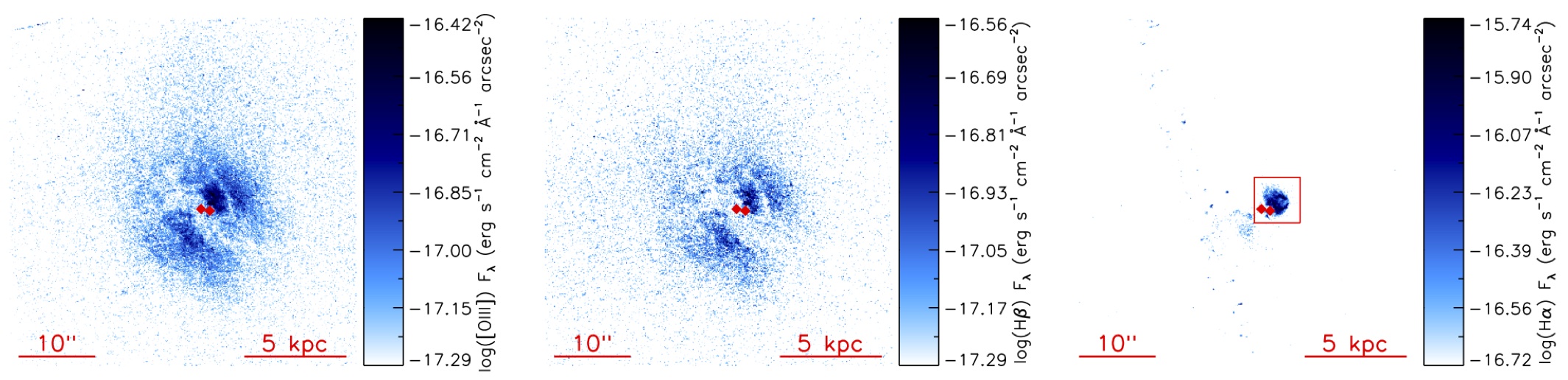}
\caption{Maps of the log([OIII]) (left), log(H$\beta$) (middle), and log(H$\alpha$+[NII]) (right) emission across the galaxy, rotated to account for a 77$^{\circ}$ P.A. so that north is up and east is left. The locations of the eastern and western nuclei, as determined in the near IR by \citet{scoville1998nicmos} and precessed to J2000.0, are shown as red diamonds. A box, showing the size of the zoomed figure~\ref{fig:o3bubmap}, is shown on the right panel. A 4$\sigma$ S/N cut has been applied. }
\label{fig:o3map}
\end{center}
\end{figure*}

\begin{figure*}[t]
\begin{center}
\includegraphics[width=\textwidth]{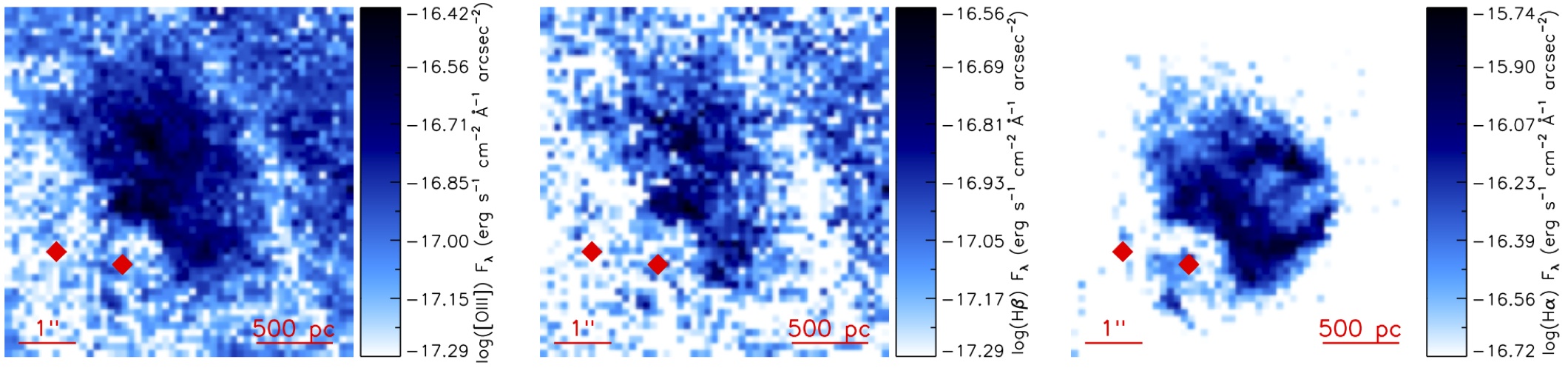}
\caption{Map of the log([OIII]) (left), log(H$\beta$) (middle), and log(H$\alpha$+[NII]) (right) emission near the nuclear region, zoomed in on a central 6'' $\times$ 6'' region. As in the full map, a 4$\sigma$ S/N cut has been applied, the locations of the eastern and western nuclei are shown as red diamonds, and the maps have been rotated so that north is up and east is left.}
\label{fig:o3bubmap}
\end{center}
\end{figure*}

\section{MORPHOLOGY}
\label{sec:morph}

\subsection{Optical emission line structure}

We present emission line maps of [OIII], H$\beta$, and H$\alpha$+[NII] in figure~\ref{fig:o3map}. Each image is 45'' $\times$ 45''. A 4$\sigma$ S/N cut has been applied, and the logarithm of the emission is displayed in order to show the full range of emission. The maps have been rotated to account for a 77$^{\circ}$ P.A., so that north is up and east is left. The NIR locations of the double nuclei \citep{scoville1998nicmos} are shown as red diamonds. Some residual cosmic rays are visible in the chip gap of the H$\alpha$+[NII] (right) panel figure~\ref{fig:o3map}, as detailed in section~\ref{sec:obs}.

The imaging reveals widespread [OIII] and H$\beta$ emission across the central disk, roughly 8 kpc in diameter. Faint tidal tails are visible to the northeast. A dust lane runs from the northeast to the southwest, obscuring the double nuclei of the system, which are only visible in the X-ray, IR, and radio. Though visible across the main disk of the system, the [OIII] and H$\beta$ emission is concentrated near the optical center, which is roughly 1.5'' offset to the northwest from the western nucleus. The H$\alpha$+[NII] emission is much more tightly concentrated near the optical center. We also present emission line maps zoomed into the nuclear region in figure~\ref{fig:o3bubmap}.

The most striking feature in the emission line maps is a ring- or bubble-shaped structure, seen in H$\alpha$+[NII] emission (figure~\ref{fig:o3bubmap}, right panel). It is unclear whether the feature is due to strong emission in [NII], H$\alpha$, or both. Strong H$\alpha$ emission indicates high rates of star formation, while high [NII] emission generally indicates the presence of shocked gas (see \citet{kewley2006metallicity}). The ring measures 1.6'' in diameter, or 600 pc at the distance of Arp 220, with a P.A. of 316$^{\circ}$ as measured from the western nucleus to the bubble center. There appears to be a faint smaller ring, inside the larger one, on the side closest to the nuclei. This is the first time the H$\alpha$ and [NII] peaks reported in \citet[their figure 1, approximately corresponding to their fibers 110--114]{arribas2001two-dimensional} have been resolved into a bubble morphology. This bubble feature and its environment are discussed further in section~\ref{ssec:multiwave}.

\subsection{Emission line ratios}
\label{ssec:ratio}

\begin{figure*}[t]
\begin{center}
\includegraphics[width=\textwidth]{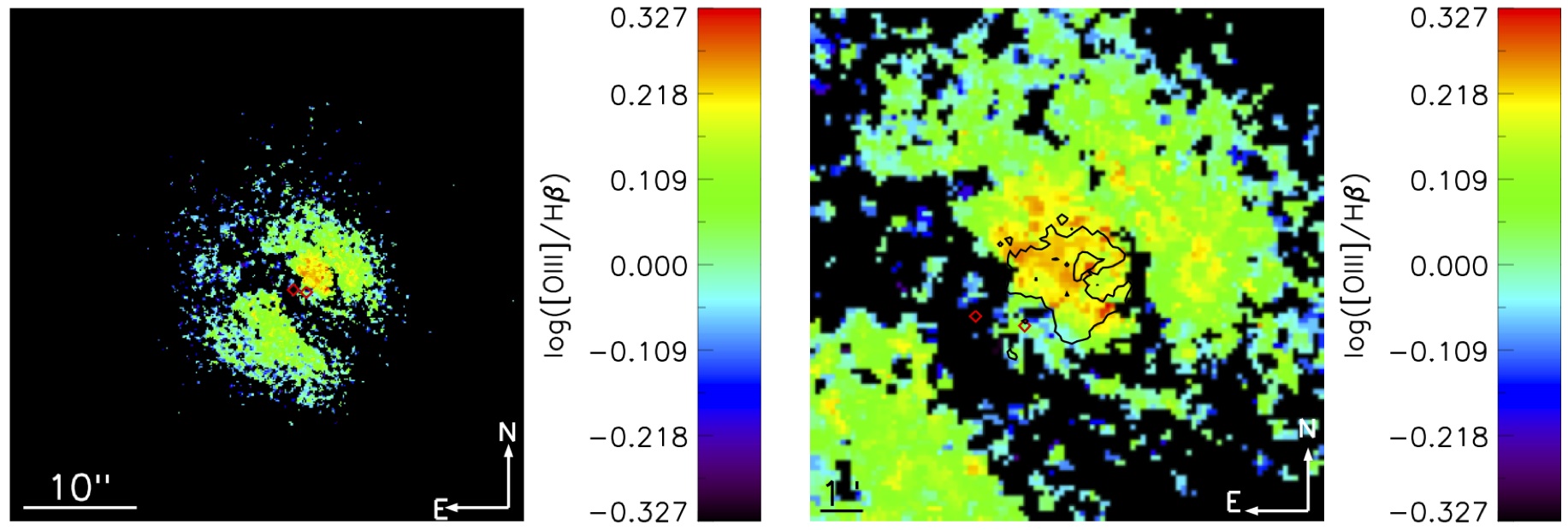}
\caption{Median smoothed maps (kernel=3) of log([OIII]/H$\beta$). Red diamonds show the positions of the nuclei, identified in the NIR and given in \citet{scoville1998nicmos}. \emph{Left:} Full view of system. \emph{Right:} Same as left, but zoomed into the optical center of the system. In addition, contours showing the bubble feature, in H$\alpha$+[NII] emission, are shown in black. North is up, east is left in both panels.}
\label{fig:o3hbmap}
\end{center}
\end{figure*}

 The ratios of strong optical emission lines are commonly used as a diagnostic to determine the excitation mechanism of ionized gas \citep{baldwin1981classification, kewley2001theoretical, kauffmann2003the-host, kewley2006the-host}: photoionization by massive stars, excitation via shocks, or ionization from an AGN. The most widely used diagnostics require the ratio of both [OIII] to H$\beta$ and either [NII], [SII], or [OI]$\lambda$ 6300\AA~to H$\alpha$.
 
In figure~\ref{fig:o3hbmap}, we present maps of log([OIII]/H$\beta$), which were created from the left and center panels of figure~\ref{fig:o3map}. A S/N cut of 4$\sigma$ was first applied to each individual emission line map and the resulting ratio map was smoothed using a median filter (kernel=3) in order to emphasize larger-scale features. The right panel of figure~\ref{fig:o3hbmap} is the same, but cropped to more closely show the region around the bubble feature. H$\alpha$+[NII] contours are overlaid in black on the right panel, to show the location of the bubble feature.

\begin{figure}[h]
\begin{center}
\includegraphics[width=\columnwidth]{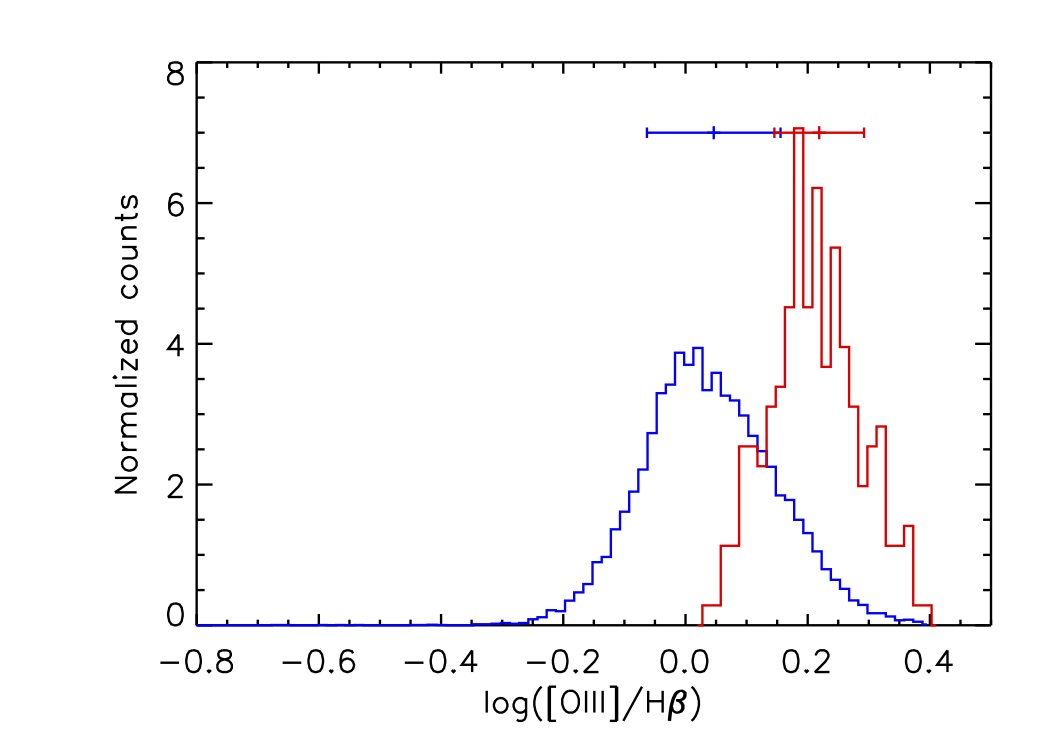}
\caption{Normalized histograms of log([OIII]/H$\beta$ values of the map of Arp 220, so that the area under each histogram is 1. The histogram in blue shows all log([OIII]/H$\beta$) values across the system (N=18676), while the histogram in red shows only the values within the H$\alpha$+[NII] bubble feature region (N=236). The mean and standard deviation of each histogram are shown at top.}
\label{fig:o3hbbubhistnorm}
\end{center}
\end{figure}

Some structure is visible in the smoothed maps, and is more readily apparent in the normalized histograms of log([OIII]/H$\beta$) shown in figure~\ref{fig:o3hbbubhistnorm}. Values for the entire system are shown in blue, while values within just the H$\alpha$+[NII] ring feature are shown in red. Binned regions are considered part of the ring feature if they fall within 1'' of the center of the ring. The mean log([OIII]/H$\beta$) for the entire system is 0.05$\pm$0.11, or 0.22$\pm$0.07 for the bubble region (shown in blue and red, respectively, above the histograms). The log([OIII]/H$\beta$) values in the ring feature appear to be slightly higher than in the system as a whole, though the difference is not significant at the 1$\sigma$ level. 

The value of log([OIII]/H$\beta$) will be used in conjunction with literature values of log([NII]/H$\alpha$) to analyze the gas excitation mechanism in section~\ref{ssec:offnucclus}. 

\subsection{The bubble in a multi-wavelength context}
\label{ssec:multiwave}

\begin{figure*}
\begin{center}
\includegraphics[width=\textwidth]{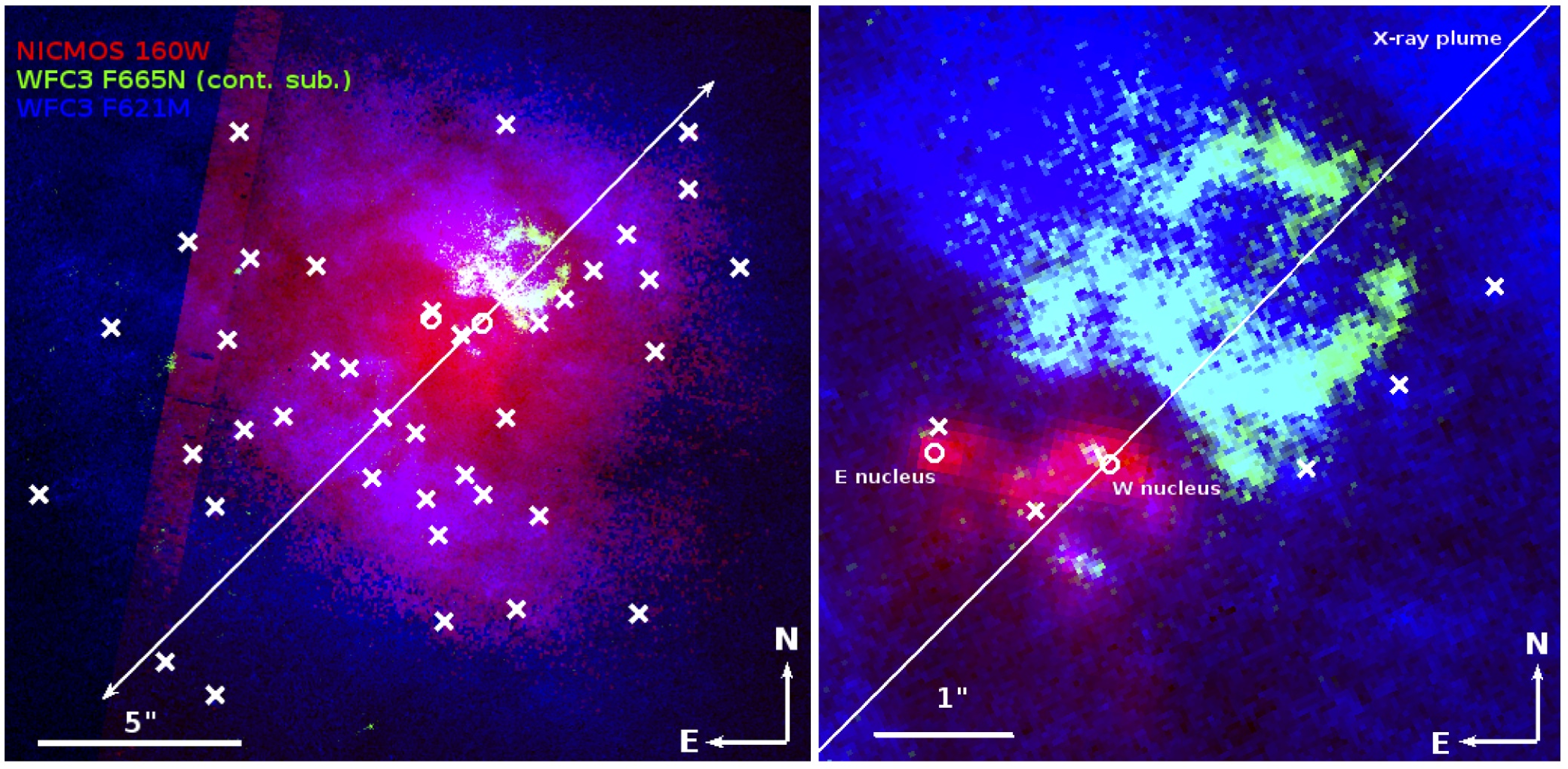}
\caption{3 color image of Arp 220. The blue image shows the optical continuum (HST/WFC3 F621M), the green shows H$\alpha$+[NII] (continuum-subtracted HST/WFC3 F665N), and the red shows the H-band infrared continuum (HST/NICMOS F160W). The major axis of the soft X-ray structure (white vector) and locations of the super star clusters (SSCs, white crosses) are indicated. The left panel shows the main disk of the system. The scaling of the red image is adjusted to emphasize the faint emission; the chip gap is visible down the left side of the image. The right image shows the nuclear region, including the double nuclei and the H$\alpha$+[NII] bubble feature. The peaks in the infrared mark the positions of the two nuclei, and are shown as white circles. The scaling of the red image has been adjusted to show the structure of the strong emission near the nuclei.}
\label{fig:3color}
\end{center}
\end{figure*}

In this section, we present the first high spatial resolution multi-wavelength maps of Arp 220 combining optical, NIR, and X-ray images. A 3-color image, shown in figure~\ref{fig:3color}, was made using the optical continuum (HST/WFC3 F621M) image, H$\alpha$+[NII] emission image, and the H-band continuum (NICMOS F160W) image. In this image, the blue shows the optical stellar continuum,  subject to extinction from dust in the system, especially in the circumnuclear region; the green shows the sum of the H$\alpha$ and [NII] emission lines, which can be excited by recent star formation, shocked gas, or AGN excitation; and the red shows a combination of the near-IR continuum, which is less affected by dust extinction.

The X-ray structure seen on $\sim$10 kpc scales, which traces the hot gas outflow, appears to be roughly aligned with the P.A. of the bubble feature with respect to the western nucleus (figure~\ref{fig:3color}). \citet[figure 2]{heckman1996rosat} reports a P.A. of 315$^{\circ}$ for the northwestern plume, while \citet[figure 1]{mcdowell2003chandra} report a P.A. of 331$^{\circ}$ for the same feature. However, the structure is wide and faint enough that determining an exact P.A. is difficult. In addition, \citet{heckman1996rosat} report a large-scale X-ray structure with a P.A. of 285$^{\circ}$, which is inconsistent with the P.A. of the bubble feature. In figure~\ref{fig:3color}, we show the mid-scale X-ray structure as reported by \citet{heckman1996rosat}. The binned pixel size in the X-ray data (1.5'') is too large to resolve whether the outflow is coming from one of the two nuclei or the $\sim$1 kpc nuclear gas disk. Higher spatial resolution observations reveal a hard X-ray source roughly aligned with the western nucleus \citep[source X-1]{clements2002chandra}. Therefore, we proceed with the assumption that the X-ray outflows originate in the western nucleus. 

The southeastern side of the bubble appears to curve around the western nucleus. This curve is echoed by the NIR emission surrounding the western nucleus, which likely traces the circumnuclear disk \citep{sakamoto1999counterrotating}. While little optical emission is seen near the nuclei, due to high dust extinction, there are peaks in the optical emission lines (H$\alpha$+[NII] and [OIII]) roughly coincident with the NIR nuclei. As the images were registered based on the coordinates of SSCs throughout the galaxy, not the nuclei, this alignment is remarkable.

A gap in the ring feature is evident on the side away from the nuclei, which could be due to either high extinction or an actual gap in emission. If the gap is caused by extinction, and the obscuring material contains hot dust, we would expect to see more NIR emission coincident with the gap. The NIR emission does not peak in this area, indicating that this hole in emission is not due to extinction from material containing hot dust. In addition, the mid-scale northwestern X-ray plume is aligned with this gap, suggesting that a stream of hot X-ray emitting gas is clearing a hole in the optical emission line gas, disrupting the process that formed these lines. This hot gas extends for 10--15 kpc on each side of the nucleus \citep{mcdowell2003chandra}.

No SSCs appear to be coincident with the bubble feature. Figure~\ref{fig:3color} shows the locations of the optical and infrared selected SSCs as white crosses \citep{wilson2006two-populations, scoville1998nicmos}. We discuss this further in section~\ref{sec:disc}.

\subsection{Kinematics from the NIR}
\label{ssec:kinNIR}

\begin{figure}
\begin{center}
\includegraphics[width=\columnwidth]{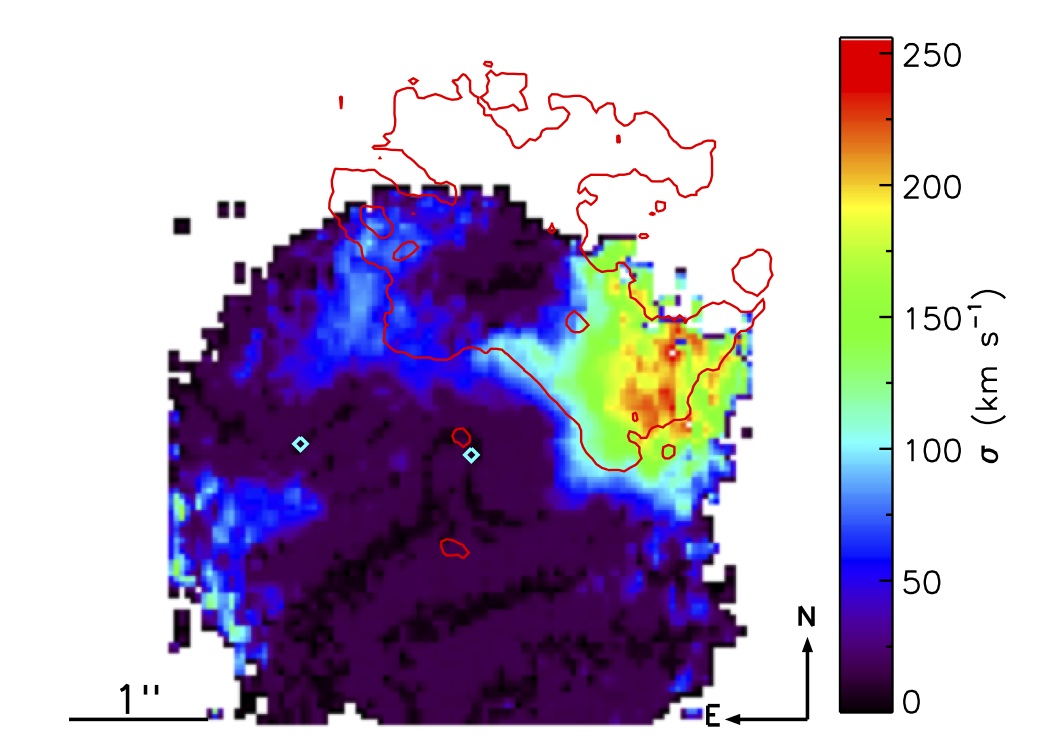}
\caption{NIR velocity dispersion, calculated from the H$_2$ 1-0 S(1) line and corrected for instrumental broadening. Contours show the 7$\sigma$ H$\alpha$+[NII] flux, smoothed for display purposes. The nuclei are marked as cyan diamonds. North is up and east is left.}
\label{fig:NIRdisp}
\end{center}
\end{figure}

\citet{engel2011arp-220:} studied the NIR kinematics of the circumnuclear region (r$\sim$1.5'', centered on the western nucleus) using the H$_2$ 1-0 S(1) line. The velocity structure presented there shows a clear rotational signature, with the rotational axis roughly in the NW/SE direction. The NIR velocity dispersion map is shown in figure~\ref{fig:NIRdisp}, with the optical H$\alpha$+[NII] flux contours overlaid. This map differs slightly from that presented by \citet{engel2011arp-220:} in that the instrumental broadening has been corrected for here; thus, the dispersion values are slightly lower than those presented earlier.  The velocity dispersion of the H$_2$ is enhanced near the bubble, and reaches a maximum of 250 km s$^{-1}$ at the base of the western arm. The velocity map shows no evidence for a systemic velocity shift at this location, though the rotational axis is not perfectly aligned with the region of enhanced velocity dispersion. This may indicate that the higher velocity dispersion is linked with the bubble feature.  If we assume an outflow is launched from a source at the center of the bubble and that the outflow velocity is equal to the peak of the dispersion, then the age of the bubble is $\sim$2.3 Myr. If the powering source is in the nucleus, a kinematic age cannot easily be determined. The current kinematic maps do not cover the bubble's full spatial extent. Additional data are necessary to characterize the dynamics of the region.

\subsection{Bubble energetics}
\label{ssec:energy}

We estimate the mass and energy in the bubble using L$_{H\alpha}$, and assuming Case B recombination and T=10$^4$ K \citep{osterbrock1989astrophysics}. Using the continuum-subtracted H$\alpha$+[NII] image, we find a total integrated flux in the bubble region of 2.6$\times$10$^{-14}$ erg s$^{-1}$ cm$^{-2}$.  With the H$\alpha$ to H$\alpha$+[NII] ratio from the bubble region of the \citet{arribas2001two-dimensional} IFS study ($\frac{H\alpha}{H\alpha+[NII]}$=0.37), we calculate an ionized gas mass of 2.5$\times$10$^7$ M$_{\odot}$ n$_e^{-1}$. 

The H$\alpha$ flux used here has not yet been corrected for extinction. We can derive the extinction in the bubble region via the Balmer decrement, by taking the ratio of the H$\alpha$ to H$\beta$ maps and comparing this value to 2.86, the theoretical ratio under Case B recombination. This calculation is very sensitive to the continuum subtraction for H$\beta$ (see section~\ref{ssec:csub}), but we can derive a range of acceptable extinction values based on the error in the continuum subtraction scaling factor. The mean continuum scaling factor for H$\beta$, 0.42, gives E(B-V)=0.11, though a range in E(B-V) of 0.00--0.48 is also consistent. Given the mean extinction, we derive a dereddened ionized gas mass of M$_{ion}$=3.2$\times$10$^7$ M$_{\odot}$ n$_e^{-1}$, with a range of 2.5--7.8$\times$10$^7$ M$_{\odot}$ n$_e^{-1}$. We adopt the mass derived from the mean value of the H$\beta$ continuum scaling factor in the following discussion.

The [SII] doublet ratio 6317\AA/6331\AA~can be used to constrain the electron density of the ionized gas. The MPA-JHU SDSS DR7 data product catalog\footnote{http://www.mpa-garching.mpg.de/SDSS/DR7/} gives a ratio of 1.46, which places n$_e$ in the lower density limit \citep{osterbrock1989astrophysics}, or less than $\sim$10$^2$ cm$^{-3}$. Previous work on the ionized bubble in NGC 3079 \citep{veilleux1994the-nuclear, cecil2001jet-} constrained n$_e$ in that system to 5--100 cm$^{-3}$. \citet{rupke2013the-multiphase} adopt n$_e$=10 cm$^{-3}$ for their analysis of the ionized winds in a sample of ULIRGs, consistent with the NGC 3079 limits. We adopt n$_e$=10 cm$^{-3}$ here to facilitate comparison with the literature. Assuming the upper limit of n$_e$=10 cm$^{-3}$ gives M$_{ion}$=3.2$\times$10$^6$ M$_{\odot}$. \citet[Table 6]{rupke2013the-multiphase} find ionized gas masses in their sample of ULIRG winds in the range 10$^{6.93-8.07}$ M$_{\odot}$, so the Arp 220 bubble mass is similar to these other ULIRG outflows.

We estimate the kinetic energy in an expanding shell by assuming that the observed velocity dispersion in H$_2$ represents the shell velocity, or v$_{shell}$=250 km s$^{-1}$. We find a kinetic energy of 2.0$\times$10$^{54}$ erg. Given a dynamical age of 2.3 Myr (section~\ref{ssec:kinNIR}) and assuming constant energy injection, we calculate an energy injection rate from the ionized gas mass of 2.8$\times$10$^{40}$ erg s$^{-1}$.

We are only observing the ionized gas and the neutral gas may be a substantial fraction of the total gas mass in the bubble feature. To include a correction for the neutral hydrogen, we used the measurements of \citet{mundell2001nuclear}, who measured hydrogen absorption against the continuum in 1.4 GHz observations. The structure they identify as spur T is coincident with the bubble feature, though they are unable to measure significant H absorption in this region. They place an upper limit of N$_H <$ 1.3$\times$10$^{20}$ T$_s$(K) cm$^{-2}$. With their assumption that T$_s$ = 100 K, that gives an upper limit of N$_H <$ 1.3 $\times$10$^{22}$ cm$^{-2}$, and an upper limit on the total neutral hydrogen mass in the bubble region of 1.2$\times$10$^8$ M$_{\odot}$. Using the same velocity assumptions as for the ionized gas, the upper limit of kinetic energy in the neutral gas is 3.8$\times$10$^{55}$ erg, and the upper limit on the energy injection rate is 5.3$\times$10$^{41}$ erg s$^{-1}$.

The total neutral+ionized mass in the bubble feature is thus M$<$1.2$\times$10$^8$ M$_{\odot}$, the total kinetic energy is KE$<$4.0$\times$10$^{55}$ erg, and the energy injection rate is dE/dt$\sim$5.6$\times$10$^{41}$ erg s$^{-1}$. We compare these rates to possible sources in the following section.

\section{ORIGIN OF THE BUBBLE}
\label{sec:disc}

Given the bubble's position and morphology, there are two possible points of origin. First, it could be the product of a jet or outflow originating in one of the two nuclei or the molecular disk immediately surrounding them. Alternatively, the bubble could originate from a source at its center. Either a massive star cluster or a microquasar (i.e. an ultraluminous X-ray source, ULX) could drive a traditional expanding superbubble. Given that the nuclei are deeply embedded in dense gas and are more distant, the energies necessary to drive an outflow from the nuclear disk versus from the bubble's center are very different. We lay out the possibilities in sections~\ref{ssec:nucorg} and \ref{ssec:offnucorg}, and discuss and compare them further in section~\ref{ssec:buborg}.

\subsection{Nuclear origins}
\label{ssec:nucorg}
The first possibility is that the bubble is being driven from a nucleus or the circumnuclear region, either from an AGN(s) or from a strong nuclear starburst.

\subsubsection{Jet or outflow from an AGN(s)}

\begin{figure}
\begin{center}
\includegraphics[width=\columnwidth]{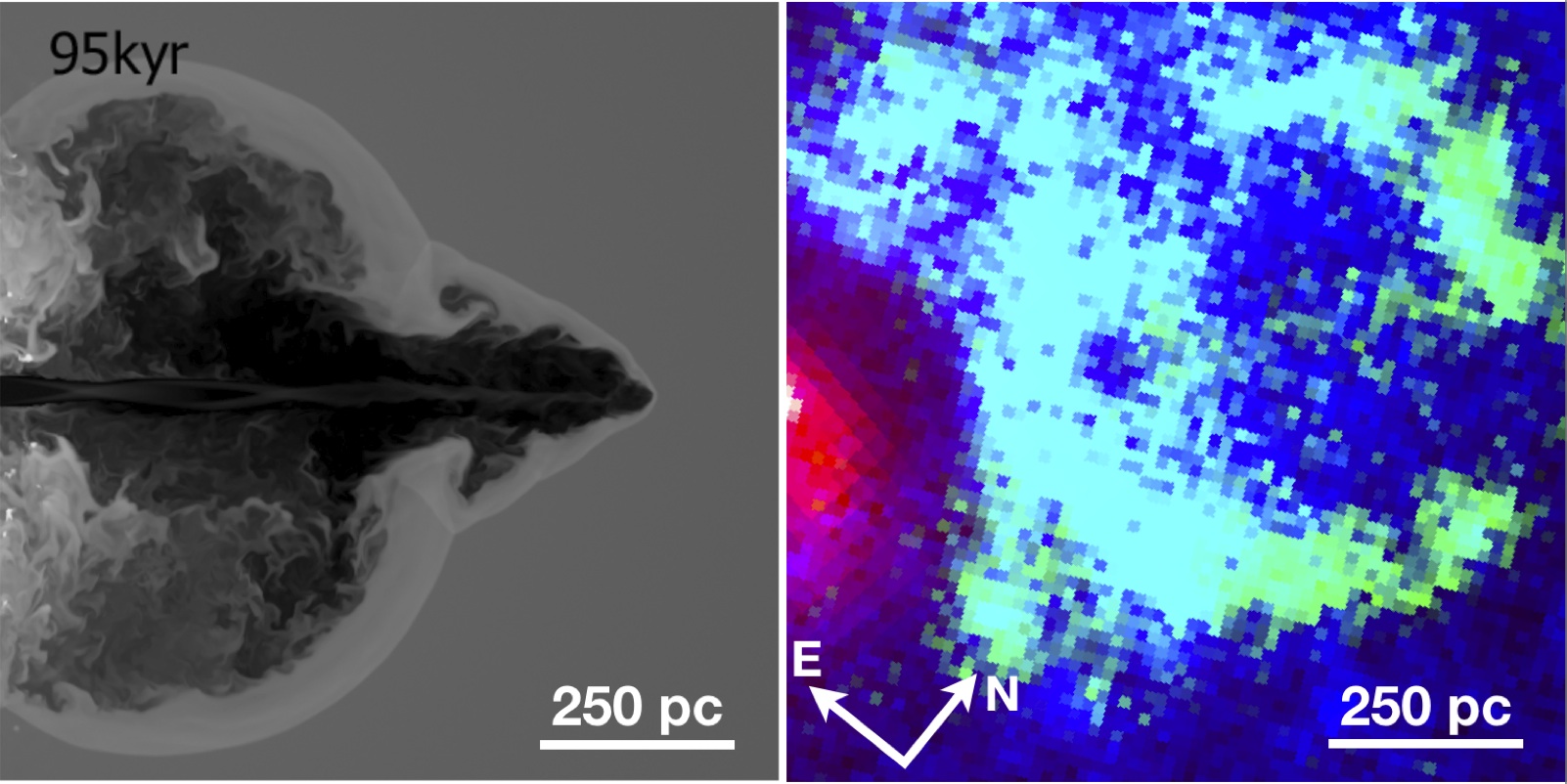}
\caption{\emph{Left:} Last panel of figure 2 from \citet{sutherland2007interactions}, reproduced here with permission. These simulations model the interaction of a supersonic jet with an inhomogenous ISM. The figure shows the 95 kyr time-step of the mid-plane density slices of the simulation, with darker colors representing less dense material. The panel is 1 kpc on a side. \emph{Right:} 3-color image of the bubble feature, as in the right panel of figure~\ref{fig:3color}, rotated and zoomed for comparison with the simulation. The western nucleus is centered on the left side of the panel and the opening in the bubble is centered on the right side of the panel. This panel is also 1 kpc on a side.}
\label{fig:AGNsim}
\end{center}
\end{figure}

The observed morphology of the bubble feature matches that produced in some AGN jet simulations.  \citet{sutherland2007interactions} modeled the interaction of a light supersonic jet with the inhomogeneous and turbulently supported ISM in an early-type galaxy. Their simulated ISM followed a lognormal single-point density distribution and a power-law density structure, with an average jet/ISM number density ratio of $\sim$10$^{-2}$. The jet had a kinetic luminosity of $\sim$10$^{43}$ erg s$^{-1}$, and radiative cooling was implemented via the MAPPINGS photoionization and shock code \citep{sutherland1993cooling}. 

The evolution of the interaction goes through several stages, including the development of a shocked bubble as the jet interacts with the ISM followed by a breakthrough of the jet through the outer boundary of the bubble. The simulation mid-plane density slices post-breakout (their figure 2, and reproduced here, in figure~\ref{fig:AGNsim}) resemble the Arp 220 bubble. However, because this is a generic jet/ISM simulation, this is an idealized comparison; in particular, the ISM of Arp 220 is likely to be  denser than that assumed here. Nonetheless, the morphological similarities suggest the possible presence of an AGN and its associated outflow. 

The proximity of the nuclei to each other (360 pc) suggests that they should be subject to strong gravitational interactions, but clear observational evidence of an AGN has been sparse. Previous X-ray studies \citep{rieke1988hard, dermer1997osse, iwasawa1999an-x-ray, iwasawa2001a-hard, clements2002chandra} ruled out a bright hard X-ray source in the nucleus, which indicates that any AGN must be highly obscured, with a lower limit on the nuclear column density of 10$^{25}$ cm$^{-2}$. However, evidence for at least one highly obscured AGN in Arp 220 has been mounting in recent years. A maser \citep{aalto2009high-resolution} and a rotating massive molecular disk  \citep{downes2007black} have been found, and \citet{contini2013the-merger} find evidence of an AGN in their modeling of the nuclear optical and infrared spectra. \citet{paggi2013two-compton-thick} find strong Fe-K$\alpha$ line emission in the vicinity of the two nuclei, concentrated most strongly on the western nucleus. From their broadband X-ray data, they estimate lower limits on the AGN bolometric luminosities of 5.2$\times$10$^{43}$ erg s$^{-1}$ (western nucleus) and 2$\times$10$^{42}$ erg s$^{-1}$ (eastern nucleus). New ALMA observations \citep{wilson2014extreme} show a luminosity surface density in excess of 10$^{14}$ L$_{\odot}$ kpc$^{-2}$ at the western nucleus, sufficiently bright to require the presence of a hot starburst, a Compton thick AGN, or both.

As shown in section~\ref{ssec:energy}, the kinetic energy input needed to form the bubble is dE/dt=5.6$\times$10$^{41}$ erg s$^{-1}$ (upper limit, ionized+neutral medium). Taking this as a ratio with respect to the bolometric luminosity of the suspected AGN in the western nucleus gives an energy deposition rate of dE/dt / L$_{AGN}$ = 1.1\%, which is an order of magnitude higher than other observed \citep{rupke2013the-multiphase} or modeled AGN energy feedback rates \citep{hopkins2010quasar}. However, these models are dependent on the fraction of gas in the hot ISM and are uncertain enough to encompass our observed value.

The evidence for an obscured AGN driving the bubble is favorable but not absolute. The morphological similarities to models and the energetics lend support to the theory that this bubble is formed by the interaction of a jet or outflow from an AGN with the galaxy's ISM. This is further supported by the large-scale X-ray structure of the galaxy, which is remarkably well aligned with not just the bubble feature, but also the gap in the side of the bubble away from the nuclei (figure~\ref{fig:3color}). However, the western nucleus (the most likely AGN host) has a gas accretion disk that is misaligned with the supposed direction of the jet and bubble. This is not definitive evidence against an AGN outflow, because the observed gas disk has a scale of 100 pc and any inner accretion disk may have a different orientation. Additionally, the outflow velocities in this system are on the low end for typical AGN \citep{rupke2005outflows3, tremonti2007the-discovery}, but may be consistent with velocities from low-luminosity AGN \citep{christopoulou1997evidence, kewley2006the-host, hicks2013fueling}. We view an AGN jet or outflow as a plausible hypothesis for this bubble's origin.

\subsubsection{Collimated outflow from a nuclear starburst}
\label{ssec:nucsb}
The prevailing explanation for the large-scale superwind structure as seen in X-rays is a massive starburst concentrated near the nuclei \citep{heckman1996rosat}. \citet{smith1998a-starburst} used VLBI continuum imaging to resolve a strong, compact OH maser near the western nucleus into a number of radio supernovae within a 75 $\times$ 150 pc region. Using the lifetime of these objects, they estimate a nuclear star formation rate of 50--800 M$_{\odot}$ yr$^{-1}$. Such high levels of star formation in the confines of the nuclear gas disk could potentially drive a starburst wind. If the starburst wind is able to break through the gas disk, the dense gas in the disk will further direct the outflow along the maximum pressure gradient. 

In section~\ref{ssec:energy}, we showed that the kinetic energy input needed to form the bubble is dE/dt=5.6$\times$10$^{41}$ erg s$^{-1}$ (upper limit, ionized+neutral medium). If we assume a SFR of 240 M$_{\odot}$ yr$^{-1}$ (equal to the global SFR and well within the nuclear SFR range), we can compare the energy in the wind to the mechanical luminosity in the nuclear starburst. Following \citet{rupke2013the-multiphase}, we calculate the mechanical luminosity in a continuous starburst using Starburst99 models from \citet{leitherer1999starburst99:}. We apply a correction for a Salpeter IMF, with a lower mass cut at 0.1 M$_{\odot}$. Thus, the energy deposition rate is dE/dt / dE/dt$_{SB99}$ = 0.8\% (upper limit). This is in line with that observed in other starburst-dominated ULIRGs \citep[3\%--13\%,][]{rupke2013the-multiphase}. It is thus also possible that the bubble feature is driven by the strong nuclear starburst.\\

\subsection{Off-nuclear origins}
\label{ssec:offnucorg}

The bubble may originate from a source at its center rather than from the nucleus. It could be the product of a massive star cluster, or of a microquasar or ULX object.

\subsubsection{Massive star cluster}
\label{ssec:offnucclus}
Structures morphologically similar to the Arp 220 bubble feature, known as superbubbles, are believed to be formed by supernovae winds from a massive star cluster pushing out into the surrounding ISM \citep{cash1980the-x-ray}. The Arp 220 bubble feature demonstrates some evidence consistent with a classical superbubble, but lacks evidence of a sufficiently massive star cluster driving the outflow.

The Arp 220 bubble shows evidence of shock excitation, which is consistent with a supernovae-driven outflow from a massive young cluster. Such outflows shock as they encounter the surrounding ISM and produce an increasing ratio of [NII] to H$\alpha$ emission at a fixed value of log([OIII]/H$\beta$) \citep{baldwin1981classification, kewley2001theoretical, kewley2006the-host}. The value of log([OIII]/H$\beta$) is fairly constant across the center of the galaxy at 0.22$\pm$0.07 (see section~\ref{ssec:ratio}, 0.05$\pm$0.11 for the galaxy as a whole). The value of log([NII]/H$\alpha$) at the maximum starburst line, or the value at which shock heating is thought to begin to dominate \citep{kewley2001theoretical}, is -0.16$\pm$0.04 for this value of log([OIII]/H$\beta)$ \citep[though there are metallicity effects; see, e.g.,][]{dopita2000a-theoretical, denicolo2002new-light, kewley2002using, kewley2006the-host}. Though we cannot separate the [NII] and H$\alpha$ emission to calculate a ratio with our data, \citet{arribas2001two-dimensional} show optical IFU spectroscopy data taken with the INTEGRAL instrument on the William Herschel telescope. Their data are lower spatial resolution (fiber diameter$\sim$0.9'', R$\sim$2000) and they identify three gas components via spectral line profile decomposition. Spatially, our bubble feature is associated with their component B. They find an average value of log([NII]/H$\alpha$) of 0.48$\pm$0.17 for this component. This is consistent with shocked gas in a shell surrounding a massive star cluster that is driving the outflow.

Although our feature resembles typical massive star cluster superbubbles, no massive star clusters are visible in the center of the bubble (figure~\ref{fig:3color}). Examination of both the optical continuum image (WFC3-UVIS F621M) and the H-band image (NICMOS F160W) reveals that the flux interior to the ring feature is high enough that a 10$^6$ M$_{\odot}$ cluster would be hidden by the background. To determine this limit, we measured the flux of other 10$^6$ M$_{\odot}$ clusters outside the bubble as identified by \citet[, clusters 2, 3, 4, 5, 11, and 12]{wilson2006two-populations}. The mean flux for a 10$^6$ M$_{\odot}$ cluster is 3.52$\pm$1.50 $\times$10$^{-18}$ ergs cm$^{-2}$ s$^{-1}$ \AA$^{-1}$ for WFC3 F621M and 1.77$\pm$0.86 $\times$10$^{-18}$ ergs cm$^{-2}$ s$^{-1}$ \AA$^{-1}$ for NICMOS F160W, using an aperture of radius 0.17''.

Analysis of the energetics provides additional evidence against an outflow driven by a massive young cluster at the bubble center. The mechanical luminosity produced by a 10$^6$ M$_{\odot}$ cluster \citep[$\sim$10$^{40}$ erg s$^{-1}$][]{leitherer1999starburst99:} is a factor of $\sim$100 below the  energy injection rate ($\sim$10$^{42}$ erg s$^{-1}$) we derived for the bubble. Thus, while we cannot rule out the existence of an undiscovered cluster smaller than 10$^6$ M$_{\odot}$, it is unlikely that it is the origin of the bubble feature.

\subsubsection{ULX bubble}

\begin{figure}
\begin{center}
\includegraphics[width=\columnwidth]{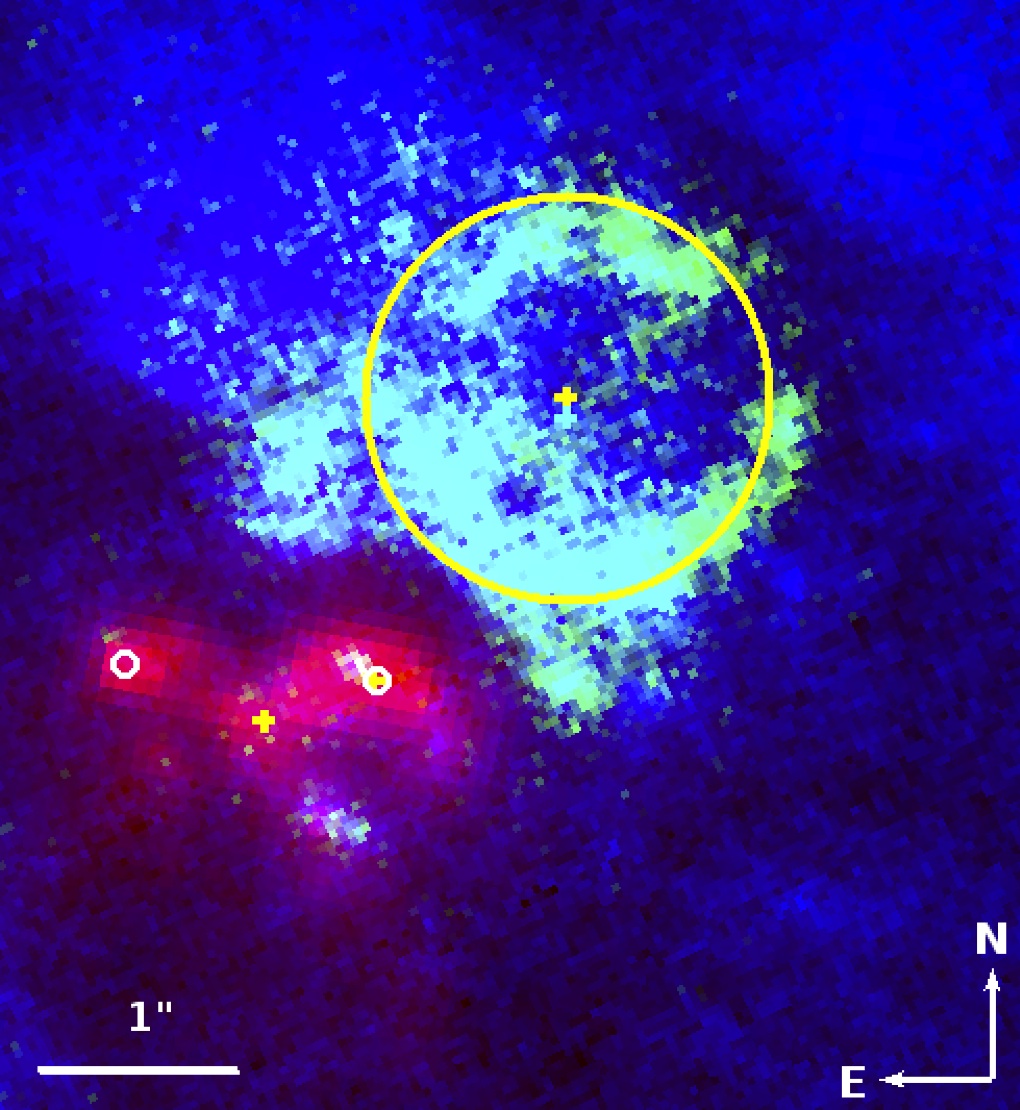}
\caption{3-color image of the central bubble. Background is the same as figure~\ref{fig:3color}, right panel. Chandra-identified X-ray sources, from \citet{clements2002chandra}, are marked as yellow crosses, and the two NIR nuclei \citep[see section~\ref{ssec:reg}]{scoville1998nicmos} are marked as white circles. From east to west, the three sources shown are their X-4, X-1 (nuclear), and X-3 (X-2 is out of the field of view). The X-ray source coordinates have been shifted so as to align their X-1 nuclear source with the western NIR nucleus; thus the western nucleus marker is obscured. X-1 and X-4 are both unresolved hard X-ray peaks, while X-3 is the peak of extended soft X-ray emission (extent of 2'', shown as the yellow circle).}
\label{fig:xray}
\end{center}
\end{figure}

Ultraluminous X-ray sources (ULX) are compact X-ray sources whose luminosities exceed the Eddington luminosity of an isotropically-emitting stellar-mass object. By definition, they are located away from the nucleus of their host galaxy. The central excitation source is still unknown, though their observed variability has been used to hypothesize these as accreting stellar- or intermediate-mass black holes \citep{colbert1999the-nature, kubota2001discovery}. They can exhibit either soft or hard X-ray spectra, and an individual ULX can transition between the two spectral modes \citep{kubota2001discovery}, presumably due to variability in accretion onto the central source. Many of these objects in the local universe (within 10 Mpc) are associated with large ionized nebulae up to several hundred pc in diameter \citep{pakull2002optical, cseh2012black, dopita2012the-physical}. These nebulae often show bright H$\alpha$ bubbles, with the X-ray source either within the bubble or on one edge. Optically, they are generally associated with strong narrow-line He II $\lambda$4686\AA~emission. This high-ionization line is not readily created by normal O-stars, though broad He II lines are observed in Wolf-Rayet stars.  

The bubble feature in Arp 220 resembles these ULX bubbles. Its physical extent, 600 pc in diameter, is roughly the same size as the known ULX bubbles. It's also coincident with a bright (L$_{0.3-10.0 keV}$ $\sim$ 10$^{39}$ ergs s$^{-1}$) soft X-ray peak observed by Chandra \citep[their X-3]{clements2002chandra} (figure~\ref{fig:xray}). However, while the X-ray sources associated with the local ULX bubbles are generally compact, Arp 220-X-3 appears to be extended, or embedded in the same extended emission that surrounds the nuclei. \citet{clements2002chandra} report an extent of 2'', shown as the yellow circle in figure~\ref{fig:xray}. No data exist on possible variability of this source, or of narrow-line He II emission. Finally, given the alignment of the bubble feature with the large-scale X-ray outflow, it is unlikely that the bubble feature has an origin unrelated to this structure. The large-scale outflow is bipolar and thus likely nuclear in origin. Further study is needed to explore whether this X-ray source is indeed a ULX with an affiliated nebula, or whether the X-ray emission is from hot gas filling the bubble created via one of the mechanisms described above.

\subsection{Bubble origins, summarized}
\label{ssec:buborg}

We have explored four possibilities for the origin of the bubble: an outflow from an obscured AGN, an outflow from high levels of star formation within the nuclear gas disk, a traditional superbubble surrounding an undiscovered star cluster at the bubble's center, or a ULX bubble. 

Given the alignment of the nuclei, bubble feature, and large-scale (biconal) outflow structure, as well as energetics arguments, it is less likely that the feature is caused by an off-nuclear source unrelated to the outflow. While a chance alignment of a ULX superbubble is possible, and warrants further study, it is more likely that an outflow or jet from the nuclear region is driving both the bubble and the outflow.

Out of the two nuclear origin scenarios, we cannot distinguish between an AGN jet or outflow or a starburst outflow. The low luminosity of the suspected AGN in the western nucleus ($\sim$10$^{43}$ erg s$^{-1}$) and the high levels of star formation within the inner nuclear gas disks around both nuclei would each provide a similar rate of energy injection, either of which would be sufficient to drive the bubble feature. It is also possible that both mechanisms are contributing to the outflow driving the bubble.

\section{CONCLUSIONS}

We have resolved the near-nuclear H$\alpha$+[NII] emission peak into a bubble feature using HST/WFC3 narrowband photometry. Our new multi-wavelength imaging maps connect this feature to both the nuclear molecular gas disk and large-scale outflows. NIR IFS data show an increase in velocity dispersion at the base of the bubble, while the optical IFS data show that the emission in this region is dominated by shock excitation. This strongly indicates that the bubble is formed by an interaction of a jet or outflow, originating in or near the nucleus, with the surrounding ISM. This outflow shocks as it propagates into the surrounding galaxy and is connected spatially with the large-scale outflow that extends for multiple kpc from the galaxy. We discuss four possibilities of the origin of the bubble and find it most plausible that the bubble originates within the inner $\sim$100 pc around the nuclei. Either an obscured AGN or the observed high level of nuclear star formation could power the observed feature. Further work is needed to differentiate between these two possibilities.

\section*{Acknowledgements}
Some of the data presented in this paper were obtained from the Mikulski Archive for Space Telescopes (MAST). STScI is operated by the Association of Universities for Research in Astronomy, Inc., under NASA contract NAS5-26555. Support for MAST for non-HST data is provided by the NASA Office of Space Science via grant NNX13AC07G and by other grants and contracts. KEL would also like to acknowledge the SWOOP writing retreat and its participants for useful feedback.

\bigskip

\section*{Appendix}
\label{sec:appa}

Narrowband imaging of [SII]$\lambda\lambda$ 6717, 6731\AA, in conjunction with similar imaging of H$\alpha$, has been used previously on dwarf galaxies \citep{calzetti2004the-ionized} to separate photoionized from non-photoionized gas, and hence to evaluate photoionized gas oxygen abundances and non-photoionized gas distributions. One of the primary motivations for obtaining the Arp 220 HST dataset presented in this paper was to apply this same technique on a late-stage merger. Unfortunately, the [SII] emission line flux could not be recovered from the F680N filter, due to both complicating factors specific to the combination of the redshift of Arp 220 and this filter, and to the shallowness of the image. In this Appendix, we describe the reduction process for this filter and suggest ways to improve the process.

The F680N filter, which contains the [SII] lines, is also contaminated by some of the H$\alpha$+[NII] line complex, which falls on the blue cutoff of the F680N bandpass (figure~\ref{fig:SDSS}). Because the bandpass cutoff falls in the middle of the intruding line complex, we cannot simply subtract the flux-calibrated H$\alpha$+[NII] image from the [SII] image. Instead, we must estimate the percentage of the contaminating lines that fall in the bandpass. We convolved the SDSS spectrum with the theoretical throughputs of the F665N and the F680N filters obtained from the \texttt{synphot} PyRAF\footnote{STSDAS and PyRAF are products of the Space Telescope Science Institute, which is operated by AURA for NASA.} package to calculate the contribution of the flux of these lines. Roughly half the flux in the H$\alpha$+[NII] line complex in the SDSS spectrum falls in the F680N bandpass, though the exact contribution depends on the width and velocity of the lines, metallicity, and gas excitation mechanism. This will likely vary across the face of the galaxy, and introduces additional uncertainties.

Changes in the line of sight velocity may shift the H$\alpha$+[NII] lines in or out of the F680N bandpass, altering the scaling factor used to remove the flux contributed by these lines. This error was estimated by red and blue shifting the SDSS spectrum by 500 km s$^{-1}$ along the line of sight. At the velocity extremes, the flux contribution from the H$\alpha$+[NII] lines complex ranges from 35\% to 80\% of the flux in these lines. The additional error in flux is included in the 1$\sigma$ errors on the [SII] image in table~\ref{tab:obs}. 

However, subtracting 50\% of the H$\alpha$+[NII] flux from the continuum-subtracted [SII] image produced oversubtracted images with negative flux values near the center of the galaxy, clear artifacts of the H$\alpha$+[NII] subtraction. This is likely due to high extinction in the central regions. We explored several avenues for improving continuum subtraction in this filter, all of which were ultimately unsuccessful at producing robust detections of [SII]. This is likely due to the intrinsic faintness of the [SII] lines, which are only 10\% of the flux in H$\alpha$+[NII] on average, and the short exposure time in the F680N filter. 

To make use of the [SII] data, we would require a longer exposure time in the F680N filter to allow us to probe the fainter average [SII] emission. Alternatively, separate high spatial resolution map of both [NII] and H$\alpha$ flux and velocity structure would allow us to more robustly correct the [SII] image for the [NII] contamination, and hence recover more [SII] flux. Without either of these, the current [SII] data are inadequate for our purposes. Future projects that attempt this type of analysis should be especially careful to obtain deep images of the fainter emission lines.

\bibliography{/Users/kel/Documents/main.bib}

\end{document}